\documentclass[a4paper,pre,10pt,amsmath,amssymb,nofootinbib,showpacs,superscriptaddress,floatfix]{revtex4}
\usepackage{amssymb}
\usepackage{graphicx}
\newcommand{\dfr}[2]{\frac {\displaystyle #1}{\displaystyle #2}}

\begin{document}

\title[]{Dynamical heterogeneity in terms of gauge theory of glass transition}

\author{Mikhail Vasin}

\address{Physical-Technical Institute, Ural Branch of Russian Academy of Sciences, 426000 Izhevsk, Russia}
\address{High Pressure Physics Institute, Russian Academy of Sciences, Moscow, Russia}
\begin{abstract}
In this paper the phenomenon of dynamic heterogeneity in supercooled liquid systems is considered in terms of the recently proposed gauge theory of glass transition. The physical interpretation of the dynamic scaling is considered. It is shown that the development of the dynamic heterogeneity occurs due to the growth areas in which molecular motion is correlated due to the elastic interaction described by the gauge field. We obtains the analytical expressions for the dependence of the number of dynamically correlated atoms as the function on the system relaxation time, and the time dependence of the dynamic susceptibility near the glass transition. It is shown that the relaxation consists two processes:  $\alpha $-relaxation process corresponding to the joint motion of the domains disordered with each other, and $\beta$-relaxation process corresponding to the motion inside these domains.
\end{abstract}

\maketitle

\section{Introduction}

The problem of the theoretical description of the liquid--glass transition is still puzzling theorists. On one hand, this transition has distinctive features of phase transition, such as critical slowing of the system, peak in the temperature dependence of susceptibility, and abrupt change of the heat capacity near the transition. On other hand, the non-equilibrium dynamics of the process determines the physical properties of the glass system to a considerable degree. For example, it leads to the dependence of the glass transition temperature on the cooling rate. The presence of these features suggests that the theory, which aspires to the full description of the liquid--glass transition, should combine both methods for describing the dynamics of nonequilibrium systems and the elements of the quasi-equilibrium theory of phase transitions. This concept has been implemented in a recently proposed gauge theory of glass transition (GTGT) \cite{Vasin}. The theory is based on the gauge approaches to glass description, which where offered at the end of the last century, \cite{Toulouse, Volovik,Hertz,Riv,Nelson,Riv1,Nusinov}, and is close ideologically to the  theory of glass transition by Kivelson \cite{Viot}.At first sight this theory seems difficult because of the necessity to use the field-theoretic technique of gauge fields. However, the results of this theory reveal fairly clear physical mechanisms of the glass transition, and allow us to clearly interpret virtually all experimentally observed properties of the glass transition.

GTGT is based on the methods of non-equilibrium dynamics, which provides a natural way to take into account the dynamic properties of the nonequilibrium vitrescent system. Therefore, it is possible to expect that this theory will allow us to move forward and describe dynamics of the glass transition.
However, as of now not all features of the dynamics of glass-forming systems have been described by this theory.
In particular, the phenomenon of the dynamic heterogeneity of the supercooled liquids has been scarcely described.
The dynamic heterogeneity is difficult to detect by direct experimental methods.
It was discovered recently in computer modeling of the glass forming liquids and became one of the most popular topics of discussion among experts in this field.
This interest is explained by the fact that the dynamic heterogeneity is considered as the expression of the fundamental mechanisms of the glass transition. Therefore, any adequate theory of glass transition should explain this phenomenon. The purpose of this paper is to discuss the phenomenon of dynamic heterogeneity in terms of GTGT.

\section{Model formulation}

Following propositions underlie GTGT:  1. It is assumed that the system is in the fluctuation region near the proposed second order phase transition, i.e. fluctuations, which represent spontaneously occurring and collapsing ordered regions, exist and amplify in the system; 2. It is assumed that the system is frustrated. The frustration, on the contrary, blocks the growth of the above-mentioned fluctuations. As shown earlier, the imposition of these conditions results in the freezing of the system in a disordered non-ergodic "solid" state, i.e. in the glass state \cite{Vasin,Vasin1}.

Because we believe that at a certain temperature $T_c$ in the system without frustration  the second order phase transition must occur, we describe the state of the ``clean'' system using the well-known Hamiltonian of the  Ginzburg-–Landau theory:
\begin{equation}\label{Action1}
    \mathcal{H}_0=\int \left[\dfr 12(\partial_i {\bf Q })^2+\dfr 12{\bf Q }^2(\mu^2 +\dfr 12v{\bf Q }^2) \right]d{\bf r},
\end{equation}
where $d{\bf r}$ denotes the volume integration, $d{\bf r}=dr_xdr_ydr_z$, $\mu^2 =\alpha (T-T_c)$, $\alpha $, and $v$ is the system parameter. In the general case one can represent tensor ${\bf Q}$ as a position-dependent orthonormal triad of unit vectors ${\bf Q}({\bf r})=[\vec Q_1({\bf r}),\,\vec Q_2({\bf r}),\,\vec Q_3({\bf r})]$ \cite{Nelson}, which is associated with the given local ordered structure, for example, with an icosahedron \cite{Nelson}.

\begin{figure}[h]
\centering
   \includegraphics[scale=0.3]{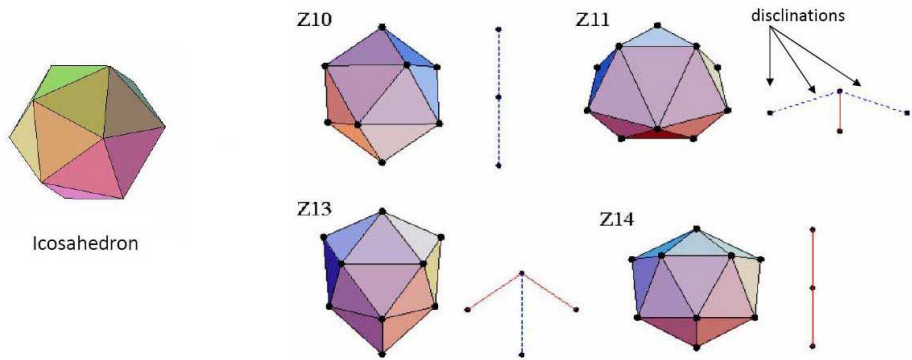}
   \caption{The simplest poly-tetrahedral model of the atomic system with the geometrical frustration is shown here.  The non-frustrated local structure of the poly-tetrahedral system is invariant with respect to the local rotations of the icosahedron symmetry group, $Y$.
   The disclinations frustrate the system, since their presence in the structure leads to the degeneracy of the configuration state space. }
   \label{Ga}
\end{figure}

The differentia of the glasses is the frustration. The frustration availability implies invariance of the system Hamiltonian with respect to local transformations, although this is not enough for the frustration of the system yet. An illustrative example is demonstrated in Fig.\,\ref{Ga}, which shows the geometrical frustration, characteristic of the dense packing of the particles with a spherically symmetric interaction potential.
In this case the local ordered state corresponds to tetrahedral packing, therefore the tensor of the local orientational order parameter is invariant with respect to the local rotations of the icosahedron symmetry group, $Y\subset$~O(3).
It turns out that the Hamiltonian (\ref{Action1}) is not invariant under such transformation, the derivative in first term hampers this. Therefore, in order to keep the gauge invariance of the continuous model, one has to move from the ordinary differentiation with respect to the spatial coordinates to the covariant differentiation: $\partial_iQ_{lk} \to D_i Q_{lk}=\partial _i Q_{lk} +g\varepsilon_{iab}A_{la}Q_{kb}$, where $\varepsilon_{iab}$ is the rotation matrices, $g$ is the topological charge, and $A_{la}$ is the gauge field which controls the rotations of $\bf Q$ in space. If $y({\bf r})$ is the matrices of the symmetry group $Y$, ${\bf Q(r)}\to y(\bf r){\bf Q(r)}$, then $\varepsilon_{iab}A_{la}({\bf r})={\bf A}({\bf r})=y({\bf r}){\bf A}({\bf r})y^{-1}({\bf r})-(\nabla y({\bf r}))y^{-1}({\bf r})$. In this case the Hamiltonian has the following gauge symmetric form \cite{Nusinov,Vasin1,VasinNCS}:
\begin{equation*}
    \mathcal{H}_0\to \mathcal{H}_{sym}=\int\left[\dfr 12(\vec D {\bf Q })^2+\dfr 12\mu^2{\bf Q }^2+\dfr 14v{\bf Q }^4)+\dfr 14{\bf F}^2\right]d{\bf r},
\end{equation*}
where $F_{a\mu\nu}=\partial_{\mu}A_{a\nu}-\partial_{\nu}A_{a\mu}+g\varepsilon_{abc} A_{b\mu}A_{c\nu}$.
It is well known that the problem of the minimisation of this Hamiltonian has got the singular solutions (vortexes) which correspond to the disclinations in the ordered atomic structure. The presence of these disclinations destroys the gauge invariance and can be described with sources ${\bf J}$ of the gauge field:
\begin{equation}\label{1}
    \mathcal{H}_{sym} \to \mathcal{H}'=\int\left[\dfr 12(\vec D {\bf Q })^2+\dfr 12\mu^2{\bf Q }^2+\dfr 14v{\bf Q }^4)+\dfr 14{\bf F}^2+{\bf JA}\right]d{\bf r},
\end{equation}
where $E_{cor}=\int {\bf JA}d{\bf r}$ is the total energy of the disclinations cores. This value should be minimal for an equilibrium system, therefore $E_{cor}\to 0$ for the ideal systems with order parameter tensor having a crystallographic symmetry. $E_{cor}\neq 0$ in the case of systems contaminated by impurities. Also $E_{cor}\neq 0$  when space can not be continually filled by the atomic structure with given symmetry of the order parameter tensor because of the topological reasons. In these cases the disclinations always present in the structure.

As it was noted above, the availability of the gauge symmetry is not enough for the frustration of the system. The system is not frustrated until the gauge field is a smooth function. Frustration corresponds to the presence a singularity of the gauge field, i.e. its source, which presents as the disclination in the atomic structure. The disclinations disorder the structure, and, in general, are mobile when the system is in the liquid state. According to the N.\,Rivier arguments \cite{Riv,Riv1} it is natural to assume that at $ T> T_c $, when the system is in the equilibrium liquid phase, the subsystem of the vortices is also in thermal equilibrium, so its temperature is $T_v=T$. Taking this into account, we can eliminate the sources of the field in (\ref {1}) by averaging over ${\bf J}$, $\langle{{\bf J}(0){\bf J}({\bf r})}\rangle=I_0({\bf r})$, where $I_0$ is a very important theory parameter, which means the equilibrium concentration of the vortices. As was noted above, in the frustrated system the disclinations always present in the structure, therefore $I_0\neq 0$, while in the non-frustrated system $I_0\to 0$. It should be noted that in the presented theory there is no principal distinction between embedded and self-induced disorder. We believe that disorder is annealed, and all information about the nature of this disorder is incorporated in parameter $I_0$. In the case of the embedded disorder, this parameter, for example, can be proportional to the number of frustrating impurities or defects entered into the system. In case of the self-induced disorder the parameter $I_0$ can depend on the number of disclinations required for continuous tiling of the Euclidean space by the elements with certain structure of local ordering, and probably can be estimated from the geometric and topological properties of this structure.

Finally, we must keep in mind that we consider the fluctuation region. According to the fluctuation theory of phase transitions, near the phase transition the order parameter can be represented as the sum of two parts: the ``slow'' part of $\bf \Phi $, and ``fast'' part of $\bf \Psi $: ${\bf Q}={\bf \Phi} + {\bf \Psi }$. ``Fast'' part contains information on the correlation functions on scales which are smaller than the correlation length, $L\ll r_c$, the ``slow'' part corresponds to the correlation functions on large scales.
The mean square of the fast part of the order parameter in the fluctuation region $|T-T_c|<T_c-T_G$ ($T_G$ is the Ginzburg temperature) will be set equal to $\langle {\bf \Psi}^2 \rangle \approx \tilde{\mu}^2/v$, where $\tilde{\mu}=\sqrt{\tilde{\alpha}(T-T_c)}$ is the renormalized value of the order parameter field ``mass''\footnote{As the renormalized value of the square ``mass'' is $\tilde{\mu}^2\approx\mu^2+v\langle{\bf \Psi}^2\rangle\gg\mu^2$ in the one-loop approximation.}.
Further, according to the fluctuation theory of phase transitions, it is necessary to eliminate the ``fast'' freedom degrees by integration of the distribution function over $\bf \Phi $.

After averaging over ${\bf J}$ and ${\bf \Psi }$ the Hamiltonian of the frustrated system model takes the final form:
\begin{equation}\label{model}
    \mathcal{H}=\int\left[\dfr 12(\nabla {\bf \Phi} )^2+\dfr{g^2}2{\bf\Phi}^2{\bf A}^2+\dfr 12M^2 {\bf A}^2+\dfr 12{\tilde{\mu}}^2 {\bf \Phi}^2 +\dfr 14v{\bf \Phi}^4+\dfr 14{\bf F}^2\right]d{\bf r}.
\end{equation}
where
\begin{equation}\label{massa}
    M^2={\tilde{\mu}}^2g^2/2v-I_0
\end{equation}
is the square of the gauge field ``mass''. From (\ref{model}) it follows that the frustrated system has another critical point, $M^2=0$, which corresponds to the new critical temperature $T_0=T_c+2I_0v/\tilde{\alpha} g^2$.

As mentioned above, GTGT is ideologically close to the relaxation mechanism proposed by Stillinger \cite{Stillinger} and to the theory of ``frustration-limited domains'' proposed by Kivelson and Tarjus \cite{Viot}, which can be obtained from the general Hamiltonian (\ref{1}) by the functional integrating not over the random sources, as in GTGT, but over the gauge field \cite{Vasin1}. However, the Kivelson approach seems to be more difficult, since the interaction between the sources (disclinations) is long-range, that leads to the breach of the condition for the Gibbs theory applicability (the energy is not additive).

\section{Critical behavior close $M^2=0$ and the scale of dynamic heterogeneity}

\begin{figure}[h]
   \centering
   \includegraphics[scale=0.35]{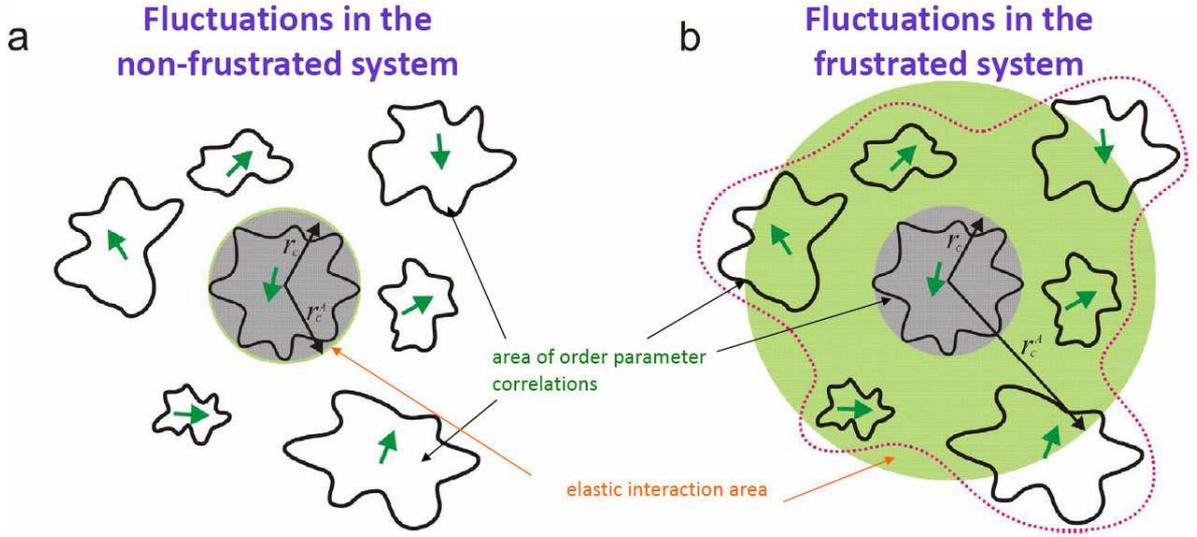}
   \caption{ Schematic representation of the fluctuations in the vicinity of the second order phase transition: a) For the non-frustrated system $r_c^A \equiv r_c$, b) in the frustrated system $r_c^A\gg r_c$. The interaction radius between the order parameter fluctuations in the frustrated system is much greater of their linear size.}
   \label{fluct}
\end{figure}

In order to describe the glass transition kinetics using GTGT one needs to study the critical behavior of the above-stated model close to the critical temperature $T_0$ (ie $ M^2=0$).
The analysis of the critical behavior of the model close to $M^2=0$, carried out in \cite{Vasin1}, has shown that at low temperatures $T\to T_c^+$
the system is in liquid state, when the average values of the fields do not change: $\langle {\bf\Phi} \rangle =0$, $\langle {\bf A} \rangle =0$,  but
 the correlation radiuses both of the gauge field, $r_c^A$, and of the order parameter field, $r_c$, grow. However, in contrast to the phase transition in the non-frustrated system in which $r_c^A \equiv r_c$, in the frustrated system the correlation length of the gauge field  grows much faster and, eventually, becomes infinitely large at $T=T_0>T_c$. At this point the correlation radius of the order parameter is finite, $r_c \ll r_c^A $, and, as a result, the described system freezes in one of the many possible states of the disordered structure. In other words, the growth of the correlation length of the order parameter is prevented by the frustration, which destroys the ordered regions, the elements of which, however, remain firmly linked.

In the quantum field theory each gauge field corresponds to a particular type of interaction. In this system only elastic interaction can be present. Hence, the physical meaning of the correlation length of the gauge field becomes transparent, this length can be naturally interpreted as the radius of the elastic interaction between the system elements. In modern literature this spatial scale is called the dynamic heterogeneity scale, $L_{het} \equiv r^A_c$, although, as one can see, here the dynamics is not so important yet.  It is only possible to indirectly observe the increase of the elastic interaction radius in form of the growth of the scale of the collective motion of detached structure elements in the outwardly disordered system. Thus, $L_{het}\sim (T-T_0)^{-\nu}$, where $\nu \approx 1/2 $.

In order to make sure that the suggested interpretation is correct below we will describe the analysis of the glass transition dynamics.

\section{Analysis of the model critical dynamics}

The dynamics of the system close to $T_0$ can be analyzed by the functional methods of non-equilibrium dynamics~\cite{Kamenev}.
In \cite{Vasin} such analysis was carried out for the system in the vicinity of the glass transition temperature $T_0$. Below omitting some details we present only the main provisions of the analysis carried out using the Keldysh technique. This technique is applicable to both classical and quantum systems. But since our problem is purely classical, the classical limit of the partition function of the system under study is as follows:
\begin{equation}\label{L2}
\displaystyle Z=\int \exp (-\mathcal{S}^*)\mathfrak{D}\vec\Phi \mathfrak{D}\vec A_{a\mu},
\end{equation}
where
\begin{equation}\label{L3}
\begin{array}{c}
\displaystyle
\mathcal{S}^*=\frac 12\int \left[\vec\Phi(t,\,{\bf r})\hat G^{-1}(t-t',\,{\bf r-r'})\vec\Phi(t',\,{\bf r'})\right. \\[12pt]
   \displaystyle \left.  +\vec A_{a\mu }(t,\,{\bf r})\hat\Delta_{\mu\nu}^{-1}(t-t',\,{\bf r-r'})\vec A_{a\nu}(t',\,{\bf r'})\right]d{\bf r}d{\bf r'}dt dt' \\[12pt]
    +\displaystyle \int\left[ g\varepsilon_{abc}(\partial_{\mu}\bar A_{a\nu })A_{b\mu}A_{c\nu}  +g\varepsilon^{abc}(\partial_{\mu}A_{a\nu })\bar A_{b\mu}A_{c\nu}+g\varepsilon_{abc}(\partial_{\mu}A_{a\nu })A_{b\mu}\bar A_{c\nu}\right.\\[12pt]
   \displaystyle \left. +g^2\varepsilon_{abc}\varepsilon_{aij}\bar A_{b\mu }A_{c\nu}A_{i\mu}A_{j\nu}+g^2 \bar A_{a\mu}A_{a\mu}\Phi^2+g^2 (A_{a\mu})^2\bar\Phi\Phi +
    v4\, \bar\Phi\Phi^3\right]d{\bf r}dt,
\end{array}
\end{equation}
 where $\vec\Phi=\left\{ \bar\Phi ,\,\Phi  \right\}$, è $\vec A_{a\mu}=\left\{ \bar A_{a\mu},\,A_{a\mu} \right\}$ are the vectors whose components are namedas ``quantum'' and ``classical'', respectively~\cite{Kamenev} (these names are conditional). $G^{-1}$ and $\Delta_{\mu\nu}^{-1}$ are the matrices, which are inverse to the corresponding matrixes of the Green's functions having the following form:
\begin{equation}\label{eq:G0-1}
    \hat G=\left(\begin{array}{rl}
                            G^K_0 & G^A_0 \\
                            G^R_0 & 0
                          \end{array}\right), \quad
    \hat \Delta_{\mu\nu}=\left(\begin{array}{rl}
                            \Delta ^K_{\mu\nu} & \Delta ^A_{\mu\nu} \\
                            \Delta ^R_{\mu\nu} & 0
                          \end{array}\right),
\end{equation}
where within the momentum representation
\begin{equation}\label{7}
   \displaystyle G^{R(A)}_0 ({\bf k},\,\omega )
   =\dfr{1}{\varepsilon _k(\mu )\pm i\Gamma_{\Phi}\omega },\qquad
   \displaystyle G^{K}_0 ({\bf k},\,\omega )=\dfr{2\Gamma_{\Phi}}{(\varepsilon _k(\mu ))^2+\Gamma_{\Phi}^2\omega^2 },
\end{equation}
$\Gamma_{\Phi}$ is the kinetic parameter of the order parameter field, and $\varepsilon _k(x)=k^2+x^2$.
At $M\to 0$
\begin{equation}\label{8}
    \displaystyle \Delta^{R(A)} _{\mu\nu}({\bf k},\,\omega )\simeq\dfr{\delta_{\mu\nu}}{k^2\pm i\Gamma_{A}\omega },\qquad
    \displaystyle \Delta^{K} _{\mu\nu}({\bf k},\,\omega )\simeq\dfr{2\Gamma_{A} \delta_{\mu\nu}}{k^4+\Gamma_{A}^2\omega^2 },
\end{equation}
where $\Gamma_A $ is the kinetic parameter of the gauge field.
\begin{figure}[h]
   \centering
   \includegraphics[scale=0.7]{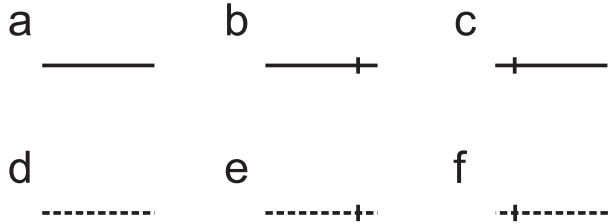}
   \caption{The graphical representation of the Green's functions: a~---~$G^K_0$, b~---~$G^A_0$, c~---~$G^R_0$, d~---~$\Delta ^K_{\mu\nu}$, e~---~$\Delta ^A_{\mu\nu}$, f~---~$\Delta ^R_{\mu\nu}$.}
   \label{fig0D}
\end{figure}
\begin{figure}[h]
   \centering
   \includegraphics[scale=0.7]{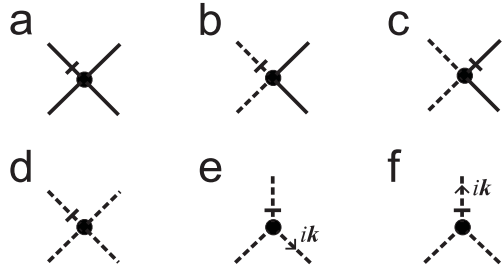}
   \caption{The graph nodes corresponding to (\ref{L3}).}
   \label{fig2D}
\end{figure}

The critical dynamics of the system in the neighborhood of $T_0$ was studied in \cite{Vasin,VasinPRB,VasinTMP}.
It was shown that at $T\to T_0^+$ the system undergoes the critical slowing down of the relaxation processes.
Since at $T \to T_0$ the interaction becomes a long-range, the system loses ergodicity.
This has a significant influence on the system kinetics and on the observed physical properties. In \cite{Vasin,VasinPRB,VasinTMP} it was shown that as a result of the ergodicity loss the relaxation time of the considered frustrated system increases with temperature decreasing according to the Vogel--Fulcher--Tamman law:
\begin{equation}\label{1N}
\tau_{\alpha} =M^2\Gamma_{\Phi }^{-1}\propto \exp\left(\dfr{\mbox{const}\cdot T_0}{T-T_0}\right),
\end{equation}
rather than a power law typical of the second order phase transitions.

\begin{figure}[h]
   \centering
   \includegraphics[scale=0.4]{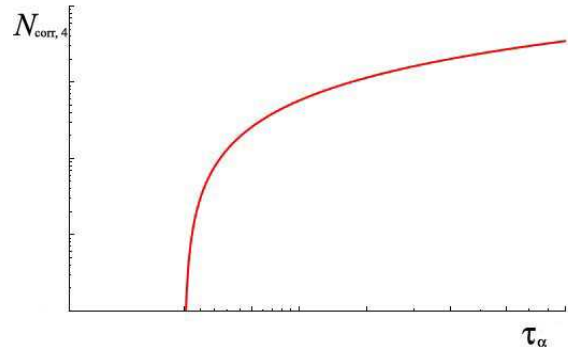}
   \caption{The qualitative form of $N_{corr, 4}(\tau_{A})$ function.}
   \label{NG}
\end{figure}

Since we assume that in the system the radius of the elastic interaction is the scale of dynamic heterogeneity, we can find the dependence of the relaxation time of the system on the number of the dynamically correlated atoms, $N_{corr, 4}\propto L_{het}^3$. For this goal it is enough to use the previously obtained expressions for $\tau_{\alpha }$ and $L_{het}$. The function $L_{het}(T)\propto (T-T_0)^{-\nu }$ is the power function. However, because $\tau_{\alpha }(T)$ is not described by a power function (\ref{1N}), then the dependence $L_{het}$ from $\tau_{\alpha }$ is not a power function either, and has the following form:
\begin{equation}\label{N}
N_{corr, 4}\propto \left(\ln \tau_{\alpha }\right)^{3/2}.
\end{equation}
This function is shown in figure \ref{NG}. One can see the characteristic slowdown of $N_{corr, 4}$ value growth with increase of the relaxation time, well known from the simulation \cite{Biroli}, which is still better interpreted as the acceleration of the growth of the system relaxation time when the temperature approaching $T_0$.

\section{Dynamic heterogeneity and dynamic susceptibility}

The conventional concepts of liquid state imply that liquid is a homogenous medium. However, on closer examination it appears that this is not so.  For example, the heterogeneity especially manifests in the supercooled liquids. The heterogeneity can have a different form. The structural heterogeneity in the mobile disordered system of atoms becomes apparent in the form of ordered domains, or the domains with different phase composition or different density. The spatial scale of these domains are quite small, even in the supercooled liquids it is a few interatomic distances. It is difficult to see how such small changes in the structure of the system can greatly affect its kinetic properties. However, recently with the aid of computer simulation  it has  become possible to confirm the old hypothesis that a much larger spatial scale has areas in which the movement of components is correlated, i.e. the movement has the cooperative character in spite of absence of the ordering. This heterogeneity is called dynamic heterogeneity \cite{Ediger}. It is exactly the advent of this dynamic heterogeneity in the system that is associated with the dramatic changes of its properties near the glass transition temperature. Below we try to describe the dynamic heterogeneity using GTGT.

In the current literature the dynamic heterogeneity of the supercooled liquids is often associated with the four-point Green function \cite{Biroli,Dasgupta,Lacevic},
\begin{equation*}
G_4({\bf r},\,t)=\langle \rho(0,\,0)\rho(0,\,t)\rho({\bf r},\,0)\rho({\bf r},\,t)\rangle - \langle \rho(0,\,0) \rho(0,\,t) \rangle \langle \rho({\bf r},\,0) \rho({\bf r},\,t) \rangle,
\end{equation*}
where $\rho ({\bf r},\,t)$ represents the density fluctuations at position ${\bf r}$ and time $t$.
It plays the same role as the standard two-point correlation function for a one-body order parameter in usual phase transitions. In particular, it is assumed that the spatial scale of the dynamic heterogeneity corresponds to the correlation radius of the four-point correlation function. It is called dynamic correlation radius. The dynamic correlation radius grows when the system approaches the glass transition point, and becomes infinitely large at $T_0$. The fact that this many-particle correlation function is a generalization of the order parameter Edwards--Anderson used in the physics of spin glasses makes the special status for this correlation function.

\begin{figure}[h]
   \centering
   \includegraphics[scale=0.7]{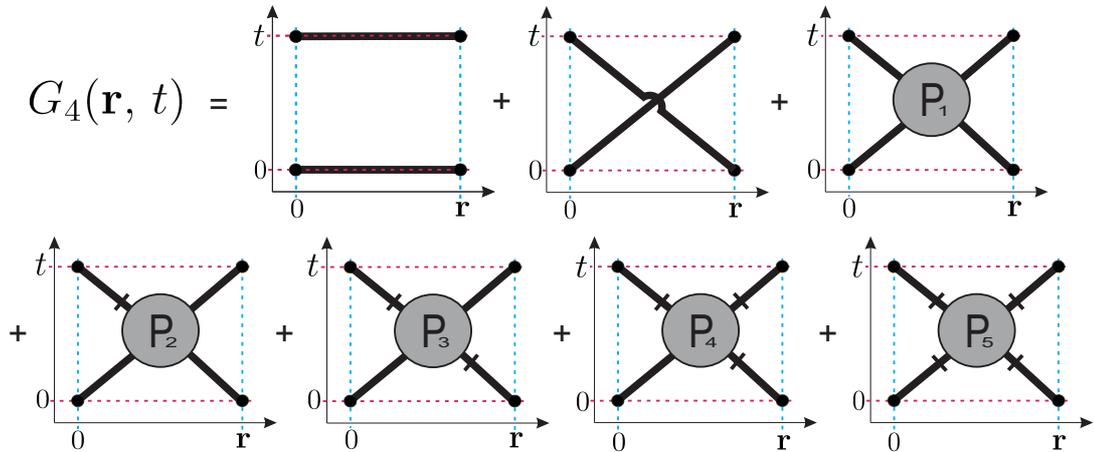}
   \caption{The graphic representation of the four-point correlation function.}
   \label{PGF}
\end{figure}

In GTGT the four-point correlation function can be calculated as the four-point Green function of the order parameter $\Phi $,
\begin{equation*}
G_4({\bf r},\,t)=\langle \Phi (0,\,0)\Phi (0,\,t)\Phi ({\bf r},\,0)\Phi ({\bf r},\,t)\rangle - \langle \Phi (0,\,0) \Phi (0,\,t) \rangle \langle \Phi ({\bf r},\,0) \Phi ({\bf r},\,t) \rangle,
\end{equation*}
since we suppose that the density fluctuations are directly connected with the ``classical'' part of the order parameter fluctuations, $\rho ({\bf r},\,t)\propto \Phi ({\bf r},\,t)$.
This function is also related to the dynamic heterogeneity, since this heterogeneity arises as a result of the correlated motion of the interconnected elements of the system.
In the graph representation it has the form presented in Fig.\,\ref{PGF}.
The vertexes $P_1,\,P_2,\dots ,\,P_5$ denote the full sets of diagrams coupling these four points.
The calculation of this function in general case is practically unsolvable problem, but at the critical point it becomes a little easier.

\begin{figure}[h]
   \centering
   \includegraphics[scale=0.8]{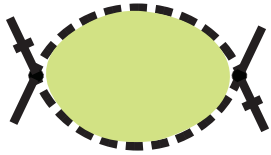}
   \caption{The loop of the gauge field Green functions, diverging close to $T_0$, is denoted by the green color.}
   \label{LOOP}
\end{figure}

\begin{figure}[h!]
   \centering
   \includegraphics[scale=0.7]{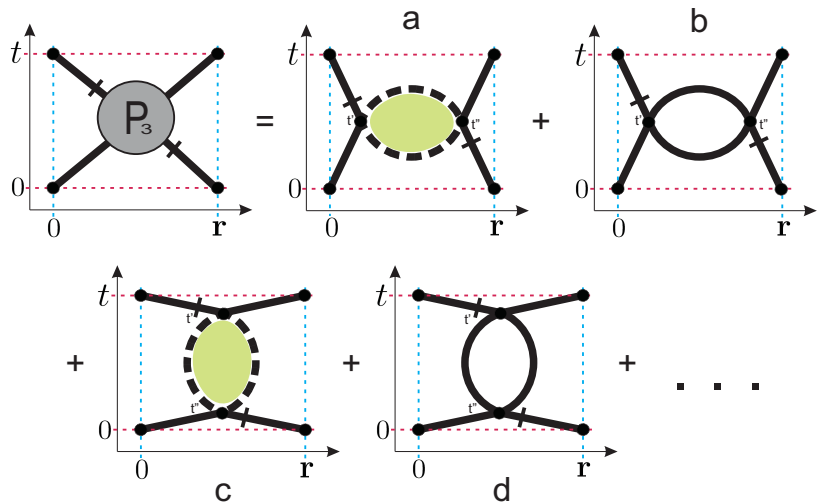}
   \caption{Here the loop of the gauge field Green functions, which contribution diverges close to $T_0$, is denoted by the green color, the non-diverging loop of the order parameter Green functions is denoted by the gray color.}
   \label{DH}
\end{figure}

Below we will consider the system which is close to the glass transition temperature, using the one-loop approximation of the theory. In this case, according to \cite{Vasin}, the dominant contribution to the correlation functions is given by the diagrams with the gauge field loop shown in Fig.\,\ref{LOOP}.
This loop plays a very important role in the description of the collective motion in glass formers. It corresponds to the exchange interaction between elements of the system, by means of the phonon corresponding to the gauge field (see Fig.\,\ref{fluct}).
In the system which is in a liquid state, the phonons have a mass and consequently, the interaction radius and the area of the correlated atomic motion are limited. In the quantum field theory this phenomenon is known as ``confinement''. At the glass transition point the phonon becomes massless, $M\to 0$, and its interaction radius diverges. Thus, the spatial scale of the dynamic heterogeneity is physically equivalent to the radius of the elastic interaction between the domains of the system, implemented by the gauge field.
For the same reason, the observed physical properties of the vitrescent system are defined by the correlation radius and the relaxation time of the gauge field. In the $({\bf k},\,t)$-representation the contribution of the loop can be described follows:
\begin{equation*}
\displaystyle \Sigma ({\bf k},\,t)=2g^4\int \dfr{\exp \left[-((k_1^2+M^2)\Gamma_A^{-1}+(({\bf k}_1+{\bf k})^2+M^2)\Gamma_{A}^{-1})|t|\right]}{(k_1^2+M^2)(({\bf k}_1+{\bf k})^2+M^2)} d{\bf k}_1,
\end{equation*}
where ${\bf k}$ is the momentum passing through the loop.
In the case of $d=3$ and $t\lesssim \Gamma_{A}M^{-2}$ it can be approximated by the following expression (see Appendix I):
\begin{equation*}
    \Sigma ({\bf k},\,t)\propto \sqrt{|t|}e^{-(k^2+M^2)|t|\Gamma_{\Phi}^{-1}} .
\end{equation*}
Let us return to Fig.\,\ref{PGF}, and note that the term $P_3$ gives the basic contribution in the four-point correlation function. It is obvious from Fig.\,\ref{DH}, where one can see that this term is determined by the gauge field loop. In the rest of the terms this loop is absent.
Close to the glass transition temperature the contributions of the terms with this loop diverges, and others dominate.

Also, the dynamic heterogeneity is usually characterized by the magnitude of the dynamic susceptibility, $\chi_4$, which is the volume integral of the four-point correlation function, $\chi_4(t)\equiv\int G_4({\bf r},\,t)d{\bf r}$ \cite{Dasgupta}.
According to GTGT the dynamic susceptibility, $\chi_4$, is related with the nonlinear susceptibility,
\begin{equation*}
\displaystyle \chi_n=\int\int \prod\limits_{i=1}^4 \left( d{\bf r}_idt_i\right) \langle \Phi({\bf r}_1,\,t_1)\bar\Phi({\bf r}_2,\,t_2)\bar\Phi({\bf r}_3,\,t_3)\bar\Phi({\bf r}_4,\,t_4)\rangle .
\end{equation*}
Both of these quantities diverge at the glass transition, since they are defined by the divergent contribution of the loops formed by the gauge field propagators.
The difference between these values is that the dynamic susceptibility is a time-dependent function, whereas the nonlinear susceptibility is the static value.
In computer simulation the dynamic susceptibility can be easily calculated, and its time dependence characterizes the dynamics of the supercooled glass-forming system. In addition, this value is the dynamic generalization of the Edwards--Anderson parameter, which is applicable to molecular glass systems. These factors make the dynamic susceptibility very popular among theorists.

\begin{figure}[h]
   \centering
   \includegraphics[scale=0.4]{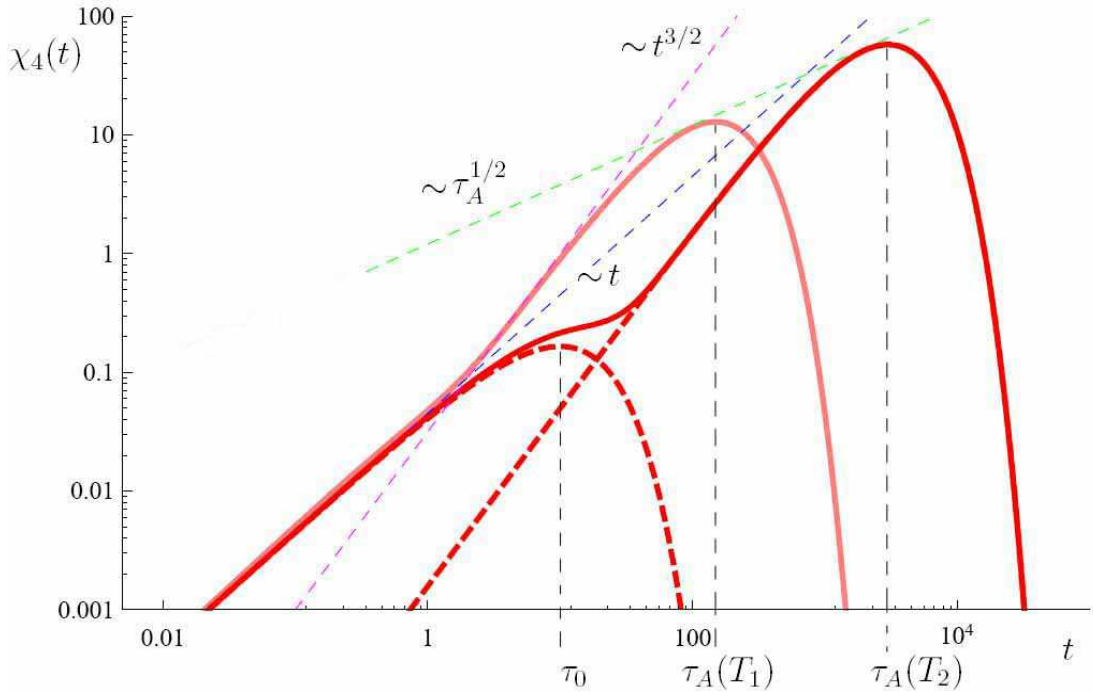}
   \caption{ The red and pink lines represent the time dependence of the dynamic susceptibility, $\chi_4$, at the temperatures $T_1$ and $T_2$, respectively ($T_1>T_2> T_0$). When $\tau_0<t<\tau_A$ the function $\chi_4(t)$ can be approximated by the power law $\chi_4(t)\propto t^{3/2}$, shown in figure by the purple dashed line. When $t<\tau_0$ the function $\chi_4(t)$ can be approximated by the power law $\chi_4(t)\propto t$, shown in figure by the blue dotted line. The maximums of the dynamic susceptibility, corresponding to the different temperatures, lay on the green dotted line corresponding to the function $\sim t^{1/2}$. The dashed red lines denote two peaks corresponding to $\alpha $- and $\beta $- relaxation processes.}
   \label{DynS}
\end{figure}

Let us see how this value can be derived within GTGT. As it was noted above, for simplicity we restrict ourselves to the one-loop approximation and take into account only the terms, which provide the main contribution to the correlation in the vicinity of the glass transition temperature. Then the diagrammatic representation of the four-point correlation function has the form shown in Fig.\,\ref{DH}, and analytical expression of the dynamic susceptibility can be written as follows:
\begin{equation*}
  \chi_4(t)\approx X_1(t)+X_2(t)+Y_1(t)+Y_2(t),
\end{equation*}
where
\begin{equation*}
\begin{array}{rl}
  X_1(t)=&\displaystyle \int d{\bf r}\int\int dt'dt''d{\bf r}'d{\bf r}'' G^K({\bf r}',\,t-t')G^R({\bf r}',\,t')\\[10pt]
  &\displaystyle \times [\Delta^K({\bf r}''-{\bf r}',\,t''-t')]^2 G^K({\bf r}-{\bf r}'',\,t-t'')G^R({\bf r}-{\bf r}'',\,t''),\\[12pt]
  X_2(t)=&\displaystyle \int d{\bf r}\int\int dt'dt''d{\bf r}'d{\bf r}'' G^K({\bf r}-{\bf r}',\,t-t')G^R({\bf r}',\,t-t')\\[10pt]
  &\displaystyle \times [\Delta^K({\bf r}''-{\bf r}',\,t''-t')]^2 G^K({\bf r}-{\bf r}'',\,t'')G^A({\bf r}'',\,t''),
\end{array}
\end{equation*}
the graphs of these contributions are shown in Fig.\,\ref{DH} ($\bf a, c$), and
\begin{equation*}
\begin{array}{rl}
  Y_1(t)&\displaystyle =\int d{\bf r}\int\int dt'dt''d{\bf r}'d{\bf r}'' G^K({\bf r}',\,t-t')G^R({\bf r}',\,t')\\[10pt]
  &\displaystyle \times [G^K({\bf r}''-{\bf r}',\,t''-t')]^2G^K({\bf r}-{\bf r}'',\,t-t'')G^R({\bf r}-{\bf r}'',\,t''),
 \\[12pt]
  Y_2(t)&\displaystyle =\int d{\bf r}\int\int dt'dt''d{\bf r}'d{\bf r}'' G^K({\bf r}-{\bf r}',\,t-t')G^R({\bf r}',\,t-t')\\[10pt]
 & \displaystyle \times [G^K({\bf r}''-{\bf r}',\,t''-t')]^2 G^K({\bf r}-{\bf r}'',\,t'')G^A({\bf r}'',\,t''),
\end{array}
\end{equation*}
the graphs of these contributions are shown in Fig.\,\ref{DH} ($\bf b, d$).
Close to the glass transition temperature the largest contribution is given by the diagrams $\bf a$ and $\bf c$, in which the loop of the gauge field propagators  diverges when $k\to 0$.
The presence of the order parameter Green functions in these diagrams leads to the exponential decrease of the contribution of these diagrams when $|t-t'|>\tilde\mu^{-2} \Gamma_{\Phi}$ and $|t''|>\tilde\mu^{-2}\Gamma_{\Phi}$.
Therefore, at the sufficiently large times, when $t\gg \tilde\mu^{-2}\Gamma_{\Phi}$, the main contribution is only given by the diagram $\bf a$ with $t'-t''\approx t$.
From this we can approximately estimate the behavior of the function $\chi_4(t) $ at large times $t$ (see. Appendix II):
\begin{equation}\label{DS1}
  \chi_4(t)\propto |t|^{3/2}\tau_A^{-1}\exp \left[-|t|/\tau_{A}\right],
\end{equation}
where $\tau_{A}=\Gamma_{A}M^{-2}\propto (T-T_0)^{-1}$ is the characteristic lifetime of the elastically coupled molecular clusters. Figure~\ref{DynS} shows the graphical form of this function in logarithmic coordinates. The analytical dependence is qualitatively in good agreement with the results of computer simulation.
The expression (\ref{DS1}) shows that the relaxation time of the four-point correlation function coincides with the relaxation time of the two-particle correlation function. It seems quite natural and coincides with the experimental data \cite{Exp1}.
At the small times $t<\tau_{A}$ the dynamic susceptibility can be approximated by the power function $\sim t^{a}$ with exponent $a=3/2$.
This dependence seems to be quite natural, since according to the dynamic susceptibility definition the response formed on the scales $\xi < L_{het}$ ($t < \tau_A$) has the form of $\chi_4\sim \xi^3$, and the dispersion relation for the diffusive process has the form of $r^2\sim t$, as a result $\chi_4\sim t^{3/2}$ on these scales.
Thus this process corresponds to the diffusive phonons motion in the disordered elastic medium.
When $t\approx \tau_A$ the obtained dependence $\chi_{4}(t)$ reaches a maximum and then decreases exponentially $\sim \exp(-t/\tau_A) $.
One can see that this peak is the so-called $\alpha $-peak. From Fig.\,\ref{DynS} one can see that the position of this peak shifts towards large times when $T\to T_0$. According to GTGT the $\tau_A$ growth is directly related to the fast growth of the characteristic linear scale, $L_{het}$,  of the disordered areas consisting of the randomly oriented ordered  but elastic-linked domains, and characterizes the increase of the system dynamic heterogeneity.
From (\ref{DS1}) one can see that the points of the dynamic susceptibility maximums lie on the power function $\chi_{4}(\tau_A)\propto |\tau_A |^{3/2}\tau_A^{- 1}\exp \left[-|\tau_A |/\tau_{A}\right] \sim \tau_A^{1/2}$.
Note that the resulting exponent, $1/2$, coincides with the exponent obtained for the three-dimensional glass system in the non-cooperative Fredrickson-Andersen model \cite{Ton}.

The diagram {\bf c} becomes essential when $t\leqslant \tau_0=\Gamma_{\Phi}/2\tilde\mu^2$. It gives the second peak at the $\chi_4(t)$ function (see Fig.\,\ref{DH}), which corresponds to the fast processes in the elastic regime and can be identified as $\beta $-peak.
When $t \ll \tau_0$ this contribution to the dynamic susceptibility increases linearly with time (see. Appendix III). In GTGT it indicates that in this time scale the susceptibility is driven by the elastic properties of the ordered domain. At $t \approx \tau_0 $ corresponding to the domain life time, this part of the dynamic susceptibility starts fall off exponentially, $\chi_4(t)\propto t\exp \left[ -t/\tau_0\right] \theta (t)$.  Unlike the $\alpha $-peak, the position of the $\beta $-peak remains virtually unchanged when $T\to T_0$, since the average size of the ordered domain grows much slower than the gauge field correlation radius.

\section{Conclusions}

Using the functional methods of nonequilibrium dynamics and the techniques of quantum field theory one can describe the characteristic features of the time dependence of the dynamic susceptibility near the glass transition temperature. This function is in qualitative agreement with the experimental results and the simulation results.
The main conclusion that can be drawn from the results of this work is that according to GTGT the appearance of the dynamic heterogeneity in the vitrescent systems is due to the growth of the areas of the correlated motion of the elastically coupled disordered groups of atoms.
GTGT allows us to interpret these groups of atoms as elastically coupled clusters, and estimate the linear size of these clusters as the correlation length of the gauge field, $A$. Thus, the dynamic heterogeneity appears due to the growth of the elastic interaction radius between system elements, surpassing the ordering scale  of these elements.

As in the previous paper \cite{Vasin} we can conclude that the full qualitative picture of the relaxation kinetics of the vitrescent system is explained by the presence two processes:  $\alpha $-relaxation process corresponding to the joint motion of the domains disordered with each other, and $\beta$-relaxation process corresponding to the motion inside these domains.
Also we managed to get the analytical expression of the dependence of the dynamically correlating atoms number on the system relaxation time, $N_{corr, 4}\propto \left(\ln \tau_{\alpha }\right)^{3/2}$, and conclude that its characteristic form is determined by the acceleration of the growth of the relaxation time of the system when approaching $T_0$.
Note that the exponent in this function coincides with the exponent of the time dependence of the dynamic susceptibility, $\chi_4(t)\sim t^{3/2}$, corresponding to the $\alpha $-relaxation processes. Actually, according to GTGT, it is the same exponent. Both these exponents and their equivalence could be checked in modelling and experiment.

\section*{Acknowledgments}
This work was supported by the RSP grant No. 14-12-01185.

\newpage

\section*{Appendix I}
Since the form of the $\Sigma(t,\,{\bf k})$ contribution is very important and has complicated form, it needs accuracy simplification in derivation.
The analytical form of this contribution is as follows:
\begin{equation*}
\displaystyle \Sigma(t,\,{\bf k})=2g^4\int \dfr{\exp \left[-((k_1^2+M^2)\Gamma_A^{-1}+(({\bf k}_1+{\bf k})^2+M^2)\Gamma_{A}^{-1})|t|\right]}{(k_1^2+M^2)(({\bf k}_1+{\bf k})^2+M^2)} d{\bf k}_1.
\end{equation*}
In the case of $d=3$ for the small time scales, $|t| \lesssim \Gamma_A M^{-2}$,
\begin{equation*}
\begin{array}{c}
\displaystyle \Sigma(t,\,{\bf k})\approx -4\pi g^4\exp \left[-k^2\Gamma_{A}^{-1}|t|\right]\int\limits_{-\infty}^{\infty}\int\limits_{-1}^{1}\dfr{\exp \left[-(2(k_1^2+M^2)-2xk_1k)\Gamma_{A}^{-1})|t|\right]}
{(k_1^2+M^2)^2} k_1^2dk_1dx\\[12pt]
\displaystyle =-8\pi g^4\exp \left[-k^2\Gamma_{A}^{-1}|t|\right]\int\limits_{-\infty}^{\infty}\dfr{\exp \left[-2(k_1^2+M^2)\Gamma_{A}^{-1}|t|\right]}
{(k_1^2+M^2)^2} k_1^2\dfr{\mbox{sh}(2k_1k\Gamma_{A}^{-1}|t|)}{2k_1k\Gamma_{A}^{-1}|t|}dk_1\\[12pt]
\displaystyle \approx -8\pi g^4\exp \left[-k^2\Gamma_{A}^{-1}|t|\right]\int\limits_{-\infty}^{\infty}\dfr{\exp \left[-2(k_1^2+M^2)\Gamma_{A}^{-1}|t|\right]}
{(k_1^2+M^2)^2} k_1^2dk_1\\[12pt]
\displaystyle = -8\pi g^4\exp \left[-(k^2+M^2)\Gamma_{A}^{-1}|t|\right]\int\limits_{-\infty}^{\infty}\left\{ \dfr{\exp \left[-2k_1^2\Gamma_{A}^{-1}|t|\right]}
{k_1^2+M^2} -\dfr{M^2\exp \left[-2k_1^2\Gamma_{A}^{-1}|t|\right]}
{(k_1^2+M^2)^2}\right\} dk_1\\[12pt]
\displaystyle =-8\pi g^4\left\{ \dfr{\pi(1+4\Gamma_{A}^{-1}M^2|t|)}
{2M}e^{-(k^2-M^2)\Gamma_{A}^{-1}|t|}\,\mbox{erfc}\left[ \sqrt{2M^2\Gamma_{A}^{-1}|t|}\right]\right.\\[12pt]
\displaystyle \left. -e^{-(k^2+M^2)\Gamma_{A}^{-1}|t|}\,\sqrt{2\pi \Gamma_{A}^{-1}|t|}\right\}.
\end{array}
\end{equation*}
Here second term dominates, since for simplification one can use the follow approximation:
\begin{equation*}
\Sigma(t,\,{\bf k})\propto 8\pi g^4\exp \left[-(k^2+M^2)\Gamma_{A}^{-1}|t|\right]\sqrt{2\pi \Gamma_{A}^{-1}|t|}.
\end{equation*}

\section*{Appendix II}
When $t$ considerably exceeds $t_1$ and $t_2$ corresponding to decay of the correlation functions $G^{A(R)}_0$ and $G^{K}_0$, and when ${\mbox{sh}(2k_1k\Gamma_{\Phi}^{-1}|t|)}\approx {2k_1k\Gamma_{\Phi}^{-1}|t|}$, the approximate integration over loops leads to
\begin{equation*}
\begin{array}{l}
\displaystyle \chi_4\approx \dfr{8\pi g^4}{\Gamma_{\Phi}^{2}}\int\int \theta(t_1)\theta(t_2)\dfr{\sqrt{2\pi \Gamma_{A}^{-1}|t|}e^{-|t|(({\bf k}_1+{\bf k}_2)^2+M^2)\Gamma_{A}^{-1}-2t_1(k_1^2+\mu^2)\Gamma_{\Phi}^{-1}-2t_2( k_2^2+\mu^2)\Gamma_{\Phi}^{-1}}}{(k_1^2+\mu^2)(k_2^2+\mu^2)} d{\bf k}_1d{\bf k}_2dt_1dt_2\\
=\displaystyle \dfr {8\pi g^4}4\int\int \dfr{\sqrt{2\pi \Gamma_{A}^{-1}|t|}e^{-|t|(({\bf k}_1+{\bf k}_2)^2+M^2)\Gamma_{A}^{-1}}}{(k_1^2+\mu^2)^2(k_2^2+\mu^2)^2} d{\bf k}_1d{\bf k}_2\\
\approx  8\pi^3 g^4\sqrt{2\pi \Gamma_{A}^{-1}|t|}e^{-|t|M^2\Gamma_{A}^{-1}}\left( \int \dfr{e^{-|t|k_1^2\Gamma_{A}^{-1}}}{(k_1^2+\mu^2)^2} k_1^2dk_1\right)^2 \\[12pt]
\displaystyle \approx 8\pi^3 g^4\sqrt{2\pi \Gamma_{A}^{-1}|t|}e^{-|t|M^2\Gamma_{A}^{-1}}\left( \sqrt{2\pi \Gamma_{A}^{-1}|t|}\right)^2\propto |t|^{3/2}e^{-|t|M^2\Gamma_{A}^{-1}}
\end{array}
\end{equation*}

\section*{Appendix III}
Here it is convenient to use the (${\bf k},\,\omega $)-representation and take into account that the main contribution to the integral is given by the part corresponding to the small $\omega $ and $k$ passing through the gauge fields loop. Therefore,
\begin{equation*}
\begin{array}{c}
\displaystyle \chi_4\approx 8\pi g^4\left[ \int \dfr{2\Gamma_{\Phi}}{((k^2+\mu^2)^2+\Gamma_{\Phi}^2\omega^2)(k^2+\mu^2+i\Gamma_{\Phi}\omega)} d{\bf k}\right]^2\\[12pt]
\displaystyle \times \int \exp \left[-M^2\Gamma_{A}^{-1}|t'|\right]\sqrt{2\pi \Gamma_{A}^{-1}|t'|} dt'.
\end{array}
\end{equation*}
After the integration over $t'$ and Fourier transformation we get
\begin{equation*}
\chi_4\approx 8\pi g^4\left[\sqrt{t}\int \dfr {\pi^{3/2}tk^2}{2^{3/2}t^2(k^2+\mu^2)^2}e^{-t(k^2+\mu^2)/\Gamma_{\Phi}}\theta(t)d(\sqrt{t}k)\right]^2\sqrt{2}\pi\Gamma_AM^{-3}.
\end{equation*}
Thus
\begin{equation*}
\displaystyle \chi_4(t)\propto te^{-t/\tau_0}\theta(t)\qquad (\tau_0 = \Gamma_{\Phi }/2\mu^2).
\end{equation*}

\end{document}